\documentclass[aps,twocolumn,showpacs,floats,prl]{revtex4}
\usepackage{graphicx}
\usepackage{dcolumn}
\usepackage{bm}
\usepackage{color}
\usepackage{amsmath}
\usepackage{subfigure}
\usepackage{enumitem}

\begin{document}

\author{L. A. Pe\~{n}a Ardila}
\author{S. Giorgini}
\affiliation{Dipartimento di Fisica, Universit\`a di Trento and CNR-INO BEC Center, I-38123 Povo, Trento, Italy}

\title{Impurity in a Bose-Einstein condensate: study of the attractive \\ and repulsive branch
using quantum Monte-Carlo methods} 

\begin{abstract} 
We investigate the properties of an impurity immersed in a dilute Bose gas at zero temperature using quantum Monte-Carlo methods. The interactions between bosons are modeled by a hard sphere potential with scattering length $a$, whereas the interactions between the impurity and the bosons are modeled by a short-range, square-well potential where both the sign and the strength of the scattering length $b$ can be varied by adjusting the well depth. We characterize the attractive and the repulsive polaron branch by calculating the binding energy and the effective mass of the impurity. Furthermore, we investigate structural properties of the bath, such as the impurity-boson contact parameter and the change of the density profile around the impurity. At the unitary limit of the impurity-boson interaction, we find that the effective mass of the impurity remains smaller than twice its bare mass, while the binding energy scales with $\hbar^2n^{2/3}/m$, where
$n$ is the density of the bath and $m$ is the common mass of the impurity and the bosons in the bath. The implications for the phase diagram of binary Bose-Bose mixtures at small concentrations are also discussed.
\end{abstract}

\maketitle

\section{I. Introduction}
\label{sec:intro}

The polaron problem is a paradigmatic topic in condensed matter physics: it concerns the effect of the quantum fluctuations of the surrounding medium on the properties of an impurity immersed in a bath. The first formulation of the problem is due to Landau and Pekar~\cite{Landau46} in their study of the motion of electrons in polar crystals. By using a variational approach valid in the limit of strong coupling it was shown that the electron becomes eventually trapped in the potential created by the self-induced deformation of the lattice. Fr\"ohlich~\cite{Froehlich54} proposed an effective Hamiltonian, which describes the coupling between charged impurities and longitudinal optical phonons of the lattice, providing the standard model description of the polaron problem. The ground-state energy of the Fr\"ohlich Hamiltonian was calculated by Feynman~\cite{Feynman55} within a variational approach based on path integrals which yields an upper bound that nicely interpolates between the weak-coupling perturbative result and the strong-coupling Landau-Pekar prediction. After a large amount of theoretical work~\cite{Mahan90} spanning more than four decades, an exact solution of the model by means of diagrammatic Monte Carlo methods was finally presented in Refs.~\cite{Prokofev98, Prokofev00}. Remarkably, both the results obtained for the polaron ground-state energy and the effective mass are in very good agreement with Feynman's findings.     

Important generalizations of the Fr\"ohlich Hamiltonian include the Holstein model on a lattice~\cite{Devreese09}, electrons coupled to acoustical phonons in a crystal~\cite{Peeters85} and, more recently, impurity particles immersed in a dilute Bose-Einstein condensed (BEC) gas~\cite{Cucchietti06, Kalas06, Jaksch08, Tempere09}.  

In the context of ultracold atoms the polaron concept has received large attention in its fermionic version, {\it i.e.} an impurity coupled to a Fermi sea~\cite{Massignan14}. Thanks to the use of a Feshbach resonance the $s$-wave scattering length between the impurity and the fermions of the bath can be tuned at will and experiments have probed various properties of attractive and repulsive Fermi polarons both in 3D~\cite{Zwierlein09, Salomon09, Grimm12} and in 2D~\cite{Koehl12}, ranging from the weak to the strong-coupling regime and including the polaron-molecule transition. 

Bose polarons, which involve a bath consisting of a BEC and therefore are directly related to the original Fr\"ohlich model, have also been realized and their dynamics experimentally investigated~\cite{Catani12, Widera12, Oberthaler13}. However, so far, there have been no studies exploiting Feshbach resonances to increase the strength of inter-species interactions, nor measuring basic polaron properties such as their binding energy, lifetime and effective mass. On the theoretical side, the self-localization of Bose polarons was investigated using mean-field approaches~\cite{Cucchietti06, Kalas06, Jaksch08, Santamore11, Timmermans13} as well as Feynman's variational method applied to the effective Hamiltonian describing the impurity~\cite{Tempere09, Novikov10}. Starting from the Fr\"ohlich Hamiltonian other studies have focused on the calculation of the radio frequency response of the polaron~\cite{Demler14}, and of its binding energy and effective mass using renormalization group~\cite{Demler14-1} and diagrammatic Monte Carlo~\cite{Devreese14} methods. A more microscopic approach based on the T-matrix approximation was used in Ref.~\cite{Rath13} where various quasiparticle properties are calculated for both attractive and repulsive Bose polarons close to a Feshbach resonance. Similar results are also obtained in Ref.~\cite{DasSarma14} by means of a variational ansatz for the wave function of the bath-impurity system.  Finally, three-body correlations were explicitly included in the theoretical treatment both at the level of perturbation theory~\cite{Bruun1} as well as within a variational approach~\cite{Bruun2}, giving rise to a significant lowering of the binding energy of attractive polarons. 

In this paper we address the problem of Bose polarons using a fully microscopic, non perturbative approach, consisting in the quantum Monte Carlo (QMC) method. This numerical technique can provide exact results for the ground-state energy and the effective mass of the impurity as a function of the parameters of the Hamiltonian describing the inter-species and intra-species interaction potentials and the density of the bosonic bath. We model these interactions using a hard-sphere potential for the inter-boson repulsion and both a purely repulsive hard-sphere and an attractive square-well potential for the impurity-boson interaction. In particular, the latter model allows one to investigate situations where the impurity-boson $s$-wave scattering length is either positive or negative giving rise to the ground-state attractive and excited-state repulsive branches of the polaron. Our analysis is limited to the case where the mass of the impurity is equal to the one of the Bose particles in the medium, but generalizations to include different mass ratios can be easily implemented within the same method. 

We investigate the properties of the Bose polaron both along the attractive and the repulsive branch. We find that for small values of the ratio $|b|/a$ of the impurity-boson to the boson-boson scattering length our results for the binding energy and the effective mass are in good agreement with second-order perturbation theory based on a Fr\"olich-like Hamiltonian describing the coupling between the impurity and the bath. At the unitary point of resonant impurity-boson scattering ($b=\pm\infty$) the binding energy is found to scale with $\hbar^2n^{2/3}/m$, where $n$ is the density of the bath and $m$ is the common mass of the particles and the impurity. We notice that this behavior is similar to the Fermi polaron case where the binding energy is proportional to the Fermi energy of the bath~\cite{Massignan14}. The effective mass ratio $m^\ast/m$ ranges from values close to one in the weak-coupling limit up to values which remain smaller than two close to the resonant point. We find no evidence of the self-localization of the polaron, which in studies based on the Fr\"ohlich model is signaled by an abrupt increase of the effective mass as the coupling strength exceeds a critical value. We believe that this wrong prediction has to do with the inadequacy of the effective Fr\"ohlich Hamiltonian in the description of the pairing mechanism which takes place close to the resonance where the impurity and one boson from the bath can form a bound state.    

We analyze the structural properties of the bosonic bath by calculating the contact parameter which characterizes the short-range behavior of the impurity-boson pair correlation function. The knowledge of how particles in the bath are distributed around the impurity enables us to evaluate the distortion of the density profile produced by the impurity.  Within the attractive square-well model, we find a pronounced peak in the density close to the impurity both on the attractive and on the repulsive branch. This peak is a result of the pairing induced by the impurity-boson potential. It
 is a short-range feature that  can not be accounted for by the Fr\"ohlich Hamiltonian which can only describe long-range distortions of the density profile. 
 
An important point to analyze is related to the existence of few-body bound states in vacuum, such as three-body Efimov-like states and deeper bound states with more than three particles. At the resonant point we calculate the energy of the deepest bound state with three and more particles ({\it i.e.} the impurity plus two or more bosons) finding evidence that such state exists only up to six particles ({\it i.e.} the impurity plus five bosons). Remarkably, the energy of these self-bound states is in absolute value much smaller than the polaron binding energy which involves the contribution from a large number of particles in the bath. One should notice that the values we obtained for the ground-state energy of the cluster states as well as the size of the largest cluster greatly depend on the details of the hard-sphere boson-boson potential used in the simulations. However, we believe that the results for the polaron binding energy at unitarity are universal and only depend on the gas parameter $na^3$ of the bath and the mass ratio between the impurity and the bosons.

The structure of the paper is as follows. In Sec.~II we first address the single-polaron problem by introducing the model Hamiltonian (subsection II-A), and by reviewing the perturbation treatment leading to the Fr\"ohlich-type Hamiltonian. Here, we also derive the results for the polaron binding energy and effective mass valid in the weak-coupling limit (subsection II-B). Finally, in subsection II-C, we briefly review the DMC method and we discuss the different trial wave functions used to describe the attractive and repulsive polaron branch. The results along the two branches concerning binding energy, effective mass, density profiles and contact parameter are presented in subsection II-D. Furthermore, subsection II-E contains a discussion of these results specific of the resonant point for the impurity-boson scattering. In Sec.~III we report on calculations of the binding energy of few-body states in vacuum at the unitary point and on the side of the resonance where a two-body bound state exists. In Sec.~IV, we generalize the problem to many impurities obeying Bose statistics and we use DMC simulations to validate the perturbative equation of state in the limit of small concentrations bearing some consequences for the phase diagram of binary mixtures. Conclusions are finally drawn in Sec.~V.

\section{II. Single impurity}
\label{sec:single-impurity}

\subsection{A. Model Hamiltonian}
We consider a system of one impurity immersed in a dilute gas of $N$ Bose particles at 
$T=0$ described by the following Hamiltonian
\begin{eqnarray}
H&=&-\frac{\hbar^2}{2m_B}\sum_{i=1}^N \nabla_i^2+\sum_{i<j}V_B(r_{ij}) 
\nonumber\\
&-&\frac{\hbar^2}{2m_I}\nabla_\alpha^2+\sum_{i=1}^NV_I(r_{i\alpha}) \;.
\label{eq:Hamiltonian}
\end{eqnarray}
Here, the first two terms represent the kinetic and the interaction energy of the bosonic bath consisting of particles of mass $m_B$ and interacting through the two-body potential $V_B$, 
which depends on the distance $r_{ij}=|{\bf r}_i-{\bf r}_j|$ between a pair of bosons. Furthermore,
$-\hbar^2/(2m_I)\nabla_\alpha^2$ is the kinetic energy of the impurity with mass $m_I$ denoted
by the coordinate vector ${\bf r}_\alpha$ and $V_I$ is the boson-impurity potential depending on
the distance $r_{i\alpha}=|{\bf r}_\alpha-{\bf r}_i|$ between the impurity and the $i$-th particle of 
the bath. The inter-boson potential $V_B$ is modeled by the hard-sphere (HS) interaction 
\begin{equation}
V_B(r)=\begin{cases}
+\infty & r< a\\
0 & r> a \;,
\end{cases}
\label{eq:Vbosons}
\end{equation}   
where the diameter $a$ coincides with the $s$-wave scattering length. The impurity-boson 
interaction, instead, is modeled by either a purely repulsive hard-sphere potential
\begin{equation}
V_I(r)=V_I^R(r)=\begin{cases}
+\infty & r< b\\
0 & r> b \;,
\end{cases}
\label{eq:Vimp1}
\end{equation}  
or an attractive square-well (SW) potential
\begin{equation}
V_I(r)=V_I^A(r)=\begin{cases}
-V_0 & r< R_0\\
0 & r> R_0 \;.
\end{cases}
\label{eq:Vimp2}
\end{equation}  
The latter is characterized by a range $R_0$ and a depth $V_0$ ($V_0>0$) chosen such as to yield the value $b$ of the scattering length. This is determined by the transcendental equation 
\begin{equation}
b=R_0\left[ 1-\frac{\tan(K_0R_0)}{K_0R_0}\right] \;,
\label{eq:Vimp3}
\end{equation}
where $K_0^2=2m_RV_0/\hbar^2$ in terms of the reduced mass $m_R=m_I m_B/(m_I+m_B)$. 
In the case of the repulsive potential $V_I^R$ the $s$-wave scattering length is always positive, whereas for the square-well potential in Eq.~(\ref{eq:Vimp3}) the value of $b$ can be either positive or negative depending on $K_0R_0$. In particular, we consider values in the range $0<K_0R_0<\pi$, corresponding to either no bound state ($K_0R_0<\pi/2$) or one bound state ($K_0R_0>\pi/2$) in the two-body sector. In this latter case the molecular binding energy $\epsilon_b$ is obtained from the 
equation 
\begin{equation}
\frac{\tan(\kappa R_0)}{\kappa R_0}=-\frac{\hbar}{R_0\sqrt{2m_R|\epsilon_b|}} \;,
\label{eq:Vimp4}
\end{equation}
where $\kappa^2=2m_R(V_0-|\epsilon_b|)/\hbar^2$. We also notice that the value $K_0R_0=\pi/2$ corresponds to the unitary limit of the impurity-boson interaction where the scattering length $b$ diverges and the binding energy $\epsilon_b$ vanishes. In the present study we consider values of the range $R_0$ that are small compared to the interboson scattering length with $a$ ranging between 5 to $20R_0$. We expect that for such a short-range potential the value $b$ of its scattering length is the only relevant parameter for all polaron properties.  
 
We restrict the analysis of the Hamiltonian (\ref{eq:Hamiltonian}) to the case where the impurity and the bosons in the bath have the same mass: $m_I=m_B=2m_R=m$. The strength of the inter-boson interactions is determined by the gas parameter $na^3$ involving the bosonic density $n=N/V$, whereas the intensity of the impurity-boson coupling is given in terms of the ratio $|b|/a$ of the two scattering lengths.

\subsection{B. Perturbation theory}
The problem of a single mobile impurity in a Bose gas can be thoroughly investigated using perturbation theory at least in the weak-coupling regime. The approach is based on the treatment 
of the bath within the Bogoliubov approximation of a dilute Bose gas described by the Hamiltonian
\begin{equation}
H_B=E_B+\sum_{\bf k} \epsilon_k \alpha_{\bf k}^\dagger\alpha_{\bf k} \;.
\label{Bogoliubov1}
\end{equation}    
Here, $E_B$ is the ground-state energy of the bosonic particles
\begin{equation}
E_B=\frac{1}{2}gnN\left(1+\frac{128}{15\sqrt{\pi}}\sqrt{na^3}\right) \;, 
\label{Bogoliubov2}
\end{equation}
where $g=\frac{4\pi\hbar^2a}{m}$ is the inter-boson coupling constant. The operators $\alpha_{\bf k}$, $\alpha_{\bf k}^\dagger$ are the annihilation/creation operators of quasiparticles related to the bosonic particle operators $a_{\bf k}$, $a_{\bf k}^\dagger$ through the standard transformations
\begin{eqnarray}
\alpha_{\bf k} &=& u_k a_{\bf k} - v_k a_{-{\bf k}}^\dagger 
\nonumber\\
\alpha_{\bf k}^\dagger &=& u_k a_{\bf k}^\dagger - v_k a_{-{\bf k}} \;,
\label{Bogoliubov3}
\end{eqnarray}  
with coefficients $u_k^2=1+v_k^2=\frac{\epsilon_k^0+gn_0+\epsilon_k}{2\epsilon_k}$ and $u_kv_k=-\frac{gn_0}{2\epsilon_k}$. The elementary excitation energies are given by the Bogoliubov spectrum
\begin{equation}
\epsilon_k=\sqrt{(\epsilon_k^0)^2+2gn_0\epsilon_k^0} \;,
\label{Bogoliubov4}
\end{equation}
where $\epsilon_k^0=\frac{\hbar^2k^2}{2m}$ is the dispersion of free particles and $n_0$ is the density of condensed particles.

At the mean-field level the interaction energy between the impurity, located in position ${\bf r}_\alpha$, and the bath is described  by the following expression
\begin{equation}
H_{\text{int}}=g_{BI}\int d{\bf r}\;n({\bf r}) \delta({\bf r}-{\bf r}_\alpha) \;,
\label{Bogoliubov5}
\end{equation}
involving the density of bosons and the interspecies coupling constant $g_{BI}=\frac{4\pi\hbar^2b}{m}$ proportional to the boson-impurity s-wave scattering length $b$. In momentum space the above interaction Hamiltonian can be recast in the form
\begin{eqnarray}
H_{\text{int}}&=&g_{BI}n
\label{Bogoliubov6}\\
&+&\frac{g_{BI}}{\sqrt{V}}\sum_{{\bf q}\neq0}e^{i{\bf q}\cdot{\bf r}_\alpha}\sqrt{n_0}(u_q+v_q)
(\alpha_{\bf q}+\alpha_{-{\bf q}}^\dagger) \;,
\nonumber
\end{eqnarray}
where the Bogoliubov approximation $\sum_{\bf k}a_{\bf k}^\dagger a_{{\bf k}+{\bf q}}\simeq \sqrt{n_0V}(u_q+v_q)(\alpha_{\bf q}+\alpha_{-{\bf q}}^\dagger)$ of the density fluctuations in terms of quasiparticle operators (\ref{Bogoliubov3}) has been used. By applying perturbation theory to the Hamiltonian $H_B+H_{\text{int}}$, as it was done for example in Ref.~\cite{Viverit02} for a Bose-Fermi mixture, one finds the following result for the ground-state energy of the system of $N$ bosons plus one impurity
\begin{equation}
E_0=E_B+gn\left( \frac{b}{a}+\frac{32}{3\sqrt{\pi}}\sqrt{na^3}\frac{b^2}{a^2}\right) \;,
\label{Bogoliubov7}
\end{equation} 
valid to order $b^2$ in the boson-impurity coupling strength~\cite{Saam69,Tempere09}.  To the same order $b^2$ one can also calculate the effective mass of the impurity obtaining the result~\cite{note1,Cucchietti06}
\begin{equation}
\frac{m^\ast}{m}=1+ \frac{64}{45\sqrt{\pi}}\sqrt{na^3}\frac{b^2}{a^2} \;.
\label{Bogoliubov8}
\end{equation}
Apart from the trivial first-order contribution $g_{BI}n$ to the ground-state energy, the second-order corrections in Eqs.~(\ref{Bogoliubov7})-(\ref{Bogoliubov8}) scale in terms of the same dimensionless parameter $\sqrt{na^3}\frac{b^2}{a^2}$, which should be much smaller than unity to ensure the validity of the perturbation approach. We also notice that the perturbative corrections to the energy and the effective mass in Eqs.~(\ref{Bogoliubov7})-(\ref{Bogoliubov8}) diverge when the inter-boson scattering length $a$ tends to zero. This feature indicates the instability of the ideal Bose gas towards clusterization around the impurity and points out the crucial role played by the repulsive interaction between the bosons.  

The Hamiltonian (\ref{Bogoliubov6}) has the general form of the Fr\"ohlich polaron Hamiltonian describing an impurity coupled to a bath of non interacting bosonic quasiparticles~\cite{Froehlich54}. This analogy was first exploited in Ref.~\cite{Cucchietti06} where the intriguing problem of the ``self-localization" of the polaron was addressed in a fashion similar to the Landau-Pekar description of electrons in ionic crystals~\cite{Landau46}. In Ref.~\cite{Tempere09} the Jensen-Feynman variational scheme is applied to both the strong and the weak-coupling regime of the effective Fr\"ohlich Hamiltonian. Both these studies predict that for $\sqrt{na^3}\frac{b^2}{a^2}\gtrsim0.7$ the impurity ``self-localizes"  inside the potential well produced by its distortion of the bosonic density. In terms of the effective mass $m^\ast$ of the impurity the self-localization phenomenon corresponds to a large enhancement of the mass ratio, $m^\ast/m\gg1$, which arises from the cloud of bosonic quasiparticles dressing the impurity. We would like to stress, however, that the Fr\"ohlich Hamiltonian $H_B+H_{\text{int}}$, where the bath and interaction term are given respectively by Eq.~(\ref{Bogoliubov1}) and (\ref{Bogoliubov6}), is an effective low-energy reduction of the microscopic Hamiltonian (\ref{eq:Hamiltonian}). Whether it captures the relevant physics of an impurity immersed in a Bose condensate when the coupling to the bath is strong is questionable and should be analyzed with care. This problem will be investigated in the remaining part of the article using quantum Monte Carlo methods that are particularly suitable to treat strongly correlated systems in a non perturbative manner.

\subsection{C. Quantum Monte Carlo method}
We use the diffusion Monte Carlo (DMC) method which aims to solve the many-body Schr\"odinger 
equation in imaginary time $\tau=it/\hbar$ for the distribution function $f({\bf R},\tau)=\psi_T({\bf R})
\Phi({\bf R},\tau)$, where $\Phi({\bf R},\tau)$ is the wave function of the system of $N$ bosons plus one impurity and $\psi_T({\bf R})$ is a trial wave function of the particle coordinates ${\bf R}=({\bf r}_1,\dots,{\bf r}_N,{\bf r}_\alpha)$ used for importance sampling. The time-dependent Schr\"odinger equation can be written as
\begin{eqnarray}
-\frac{\partial f({\bf R},\tau)}{\partial\tau}&=& -D\nabla_{\bf R}^2 f({\bf R},\tau)+D\nabla_{\bf R}\cdot
\left[ {\bf F}({\bf R})f({\bf R},\tau)\right]
\nonumber\\
&+& \left[E_L({\bf R})-E_0\right] f({\bf R},\tau) \;,
\label{eq:TDequation1}
\end{eqnarray}
in terms of the so called local energy $E_L({\bf R})=(H\psi_T({\bf R})/\psi_T({\bf R})$ and quantum drift force ${\bf F}({\bf R})=(2\nabla_{\bf R}\psi_T({\bf R}))/\psi_T({\bf R})$. In the above equation $D=\hbar^2/(2m)$ plays the role of a diffusion constant and $E_0$ is a reference energy. The formal solution of the equation is given by
\begin{equation}
f({\bf R}^\prime,\tau+\delta\tau)=\int d{\bf R}\; G({\bf R}^\prime,{\bf R},\delta\tau) f({\bf R},\tau) \;,
\label{eq:TDequation2}
\end{equation}
where one introduces the Green's function $G({\bf R}^\prime,{\bf R},\delta\tau)=\langle{\bf R}^\prime|e^{-\delta\tau\,{\cal L}}|{\bf R}\rangle$ describing the time evolution governed by the Langevin operator ${\cal L}\equiv-D\nabla_{\bf R}^2+D\nabla_{\bf R}\cdot{\bf F}+(E_L-E_0)$. If the short-time dependence of $G({\bf R}^\prime,{\bf R},\delta\tau)$ is known for sufficiently small $\delta\tau$, the asymptotic solution for large times, $f({\bf R},\tau\to\infty)$, can be obtained by iterating Eq.~(\ref{eq:TDequation2}) for a large number of time steps $\delta\tau$. On general grounds, the initial time distribution $f({\bf R},0)=\psi_T({\bf R})\sum_n c_n\Phi_n({\bf R})$ is expanded in terms of the eigenfunctions $\Phi_n({\bf R})$ of the system corresponding to the eigenenergies $E_n$. For a system of bosons, provided the coefficient $c_0$ does not vanish, the solution at large times of Eq.~(\ref{eq:TDequation1}) is given by the expression
\begin{equation}
f({\bf R},\tau\to\infty)=c_0\psi_T({\bf R})\Phi_0({\bf R})\;,
\label{eq:TDequation3}
\end{equation}
and is proportional to the nodeless ground state $\Phi_0({\bf R})$ with energy $E_0$. The value of the ground-state energy is determined from the condition of keeping the probability distribution 
$f({\bf R},\tau)$ stationary at large times or, more conveniently, from the average of the local energy $E_L({\bf R})$
\begin{equation}
E_0=\frac{\int d{\bf R}\;f({\bf R},\tau\to\infty)E_L({\bf R})}{\int d{\bf R}\;f({\bf R},\tau\to\infty)} \;.
\label{localenergy}
\end{equation}
Apart from statistical errors, the DMC method allows one to calculate the exact ground-state energy of a system of Bose particles. Importantly, the energy estimate obtained using the DMC technique with importance sampling is to a large extent independent of the detailed shape of the trial wave function as long as $\psi_T({\bf R})$ is positive definite. If the function $\psi_T({\bf R})$ changes sign in some regions of the configuration space, the DMC algorithm guided by this trial function yields, instead of the ground state, the lowest-energy eigenstate compatible with the nodal constraint fixed by $\psi_T$.

The general form of the trial wave function used in the present study is given by
\begin{equation}
\psi_T({\bf R})=\prod_{i=1}^N f_I(r_{i\alpha})\prod_{i<j}f_B(r_{ij}) \;,
\label{trial1}
\end{equation}
where the functions $f_B$ and $f_I$  describe, respectively, inter-boson and impurity-boson two-body correlations. The functional form of the inter-boson term is constructed from the two-body scattering solution of the hard-sphere potential in Eq.~(\ref{eq:Vbosons})
\begin{eqnarray}
f_B(r)=\begin{cases}
0 & r< a\\
\frac{\sin [k(r-a)]}{r} & a<r<L/2 \;,
\end{cases}
\label{Jastrow1}
\end{eqnarray} 
where the value of the wave vector $k$ is chosen such that the first derivative of the function vanishes at half of the size $L=V^{1/3}$ of the cubic simulation box: $f_B^\prime(r=L/2)=0$. This condition ensures that the Jastrow factor (\ref{Jastrow1}) is compatible with the periodic boundary conditions used in the simulation.

For the impurity-boson correlation function $f_I$ we use instead different forms depending on the type of potential $V_I$, hard sphere or square well, and on the polaron branch, repulsive or attractive. 
\begin{itemize}[leftmargin=*]
\item{Hard-sphere potential:} We use the two-body scattering solution of the potential (\ref{eq:Vimp1})  
\begin{eqnarray}
f_I(r)=\begin{cases}
0 & r< b\\
\frac{\sin [k(r-b)]}{r} & b<r<L/2 \;,
\end{cases}
\label{Jastrow2}
\end{eqnarray} 
similar to Eq.~(\ref{Jastrow1}) with the only difference that the value of the scattering length is now given by $b$. 
\item{Square-well potential:} We use different functional forms of $f_I$ for the repulsive and attractive branch of the polaron.
\end{itemize}
\begin{enumerate}[leftmargin=*]
\item{Repulsive branch ($b>0$):} 
$f_I$ is constructed from the zero-energy scattering solution of the potential (\ref{eq:Vimp2}), orthogonal to the bound state existing at $b>0$ for two particles:
\begin{eqnarray}
f_I(r)=\begin{cases}
A\frac{\sin(K_0r)}{r} & r< R_0\\
1-\frac{b}{r} & R_0<r<\bar{R} \\
B+C\left(e^{-\alpha r}-e^{-\alpha(L-r)}\right) & \bar{R}<r<L/2\;.
\end{cases}
\label{Jastrow3}
\end{eqnarray} 
Here $\bar{R}$ is a matching point and for $r>\bar{R}$ the function $f_I$ goes to a constant reached at $L/2$ where $f_I^\prime(r=L/2)=0$. The coefficients $A$, $B$ and $C$ ensure the continuity of $f_I$ and of its first derivative at the points $R_0$ and $\bar{R}$.
\item{Attractive branch ($b<0$):}
$f_I$ is constructed in the same way as for the repulsive branch of Eq.~(\ref{Jastrow3}), with the only difference that the scattering length $b$ is negative.
\item{Attractive branch ($b>0$):} We use the solution of the two-body bound state with energy 
$\epsilon_b$ given by Eq.~(\ref{eq:Vimp4})
\begin{eqnarray}
f_I(r)=\begin{cases}
A\frac{\sin(\kappa r)}{r} & r< R_0\\
\frac{e^{-\sqrt{m|\epsilon_b|}r/\hbar}}{r}& R_0<r<\bar{R} \\
B+C\left(e^{-\alpha r}+e^{-\alpha(L-r)}\right) & \bar{R}<r<L/2\;,
\end{cases}
\label{Jastrow4}
\end{eqnarray} 
where $\kappa^2=m(V_0-|\epsilon_b|)/\hbar^2$. The coefficients $A$, $B$ and $C$ again ensure the continuity of $f_I$ and of its first derivative at the potential range $R_0$ and at the matching point $\bar{R}$.
\end{enumerate}

In all the above three cases, the values of the matching point $\bar{R}$ and of the parameter $\alpha$ are optimized by minimizing the variational energy. We notice that the function $f_I$ is positive definite along the attractive branch, whereas it changes sign at the value $r=b$ on the repulsive branch. For positive values of $b$ the nodal surface in the many-body trial wave function $\psi_T({\bf R})$ which originates from the choice (\ref{Jastrow3}) of the Jastrow correlation term allows one to discriminate between the ground-state attractive branch and the excited-state repulsive branch. 

Furthermore, we notice that the unitary limit, corresponding to $b=\pm\infty$, is reached following the attractive branch. This limit corresponds to a Jastrow term $f_I(r)\propto 1/r$ in the range $R_0<r<\bar{R}$ and is obtained by approaching the resonance both from $b>0$ and from $b<0$.

\subsection{D. Attractive and repulsive polaron branch}

Simulations are carried out using periodic boundary conditions and the number $N$ of bosons in the bath is typically $N=64$. Calculations with different numbers of particles up to $N=128$ are also performed in order to check that finite-size effects are below statistical uncertainty.
  
\begin{itemize}[leftmargin=*]
\item{\underline{Binding energy}}
\end{itemize}

We determine the polaron binding energy by calculating the energy difference
\begin{equation}
\mu=E(N,1)-E_0(N) \;,
\label{polaron-energy}
\end{equation}
where $E_0(N)$ is the ground-state energy of the system of $N$ bosons alone and $E(N,1)$ is the energy of the system of $N$ bosons plus the impurity in the same volume $V$. 

Two different branches are obtained depending on the impurity-boson interaction potential and on the choice of the Jastrow term $f_I$ in the trial function (\ref{trial1}). The attractive branch, which corresponds to $E(N,1)$ being the ground state of the composite system, is simulated using the square-well potential (\ref{eq:Vimp2}) and the positive definite function $f_I$ described in Sec.~II~C. Along the repulsive branch $E(N,1)$ is still the ground state of the hard-sphere potential (\ref{eq:Vimp1}), but it corresponds to an excited state of the potential (\ref{eq:Vimp2}) which we calculate by imposing the nodal constraint given by Eq.~(\ref{Jastrow3}) on the trial function.

\begin{figure}
\begin{center}
\includegraphics[width=8.0cm]{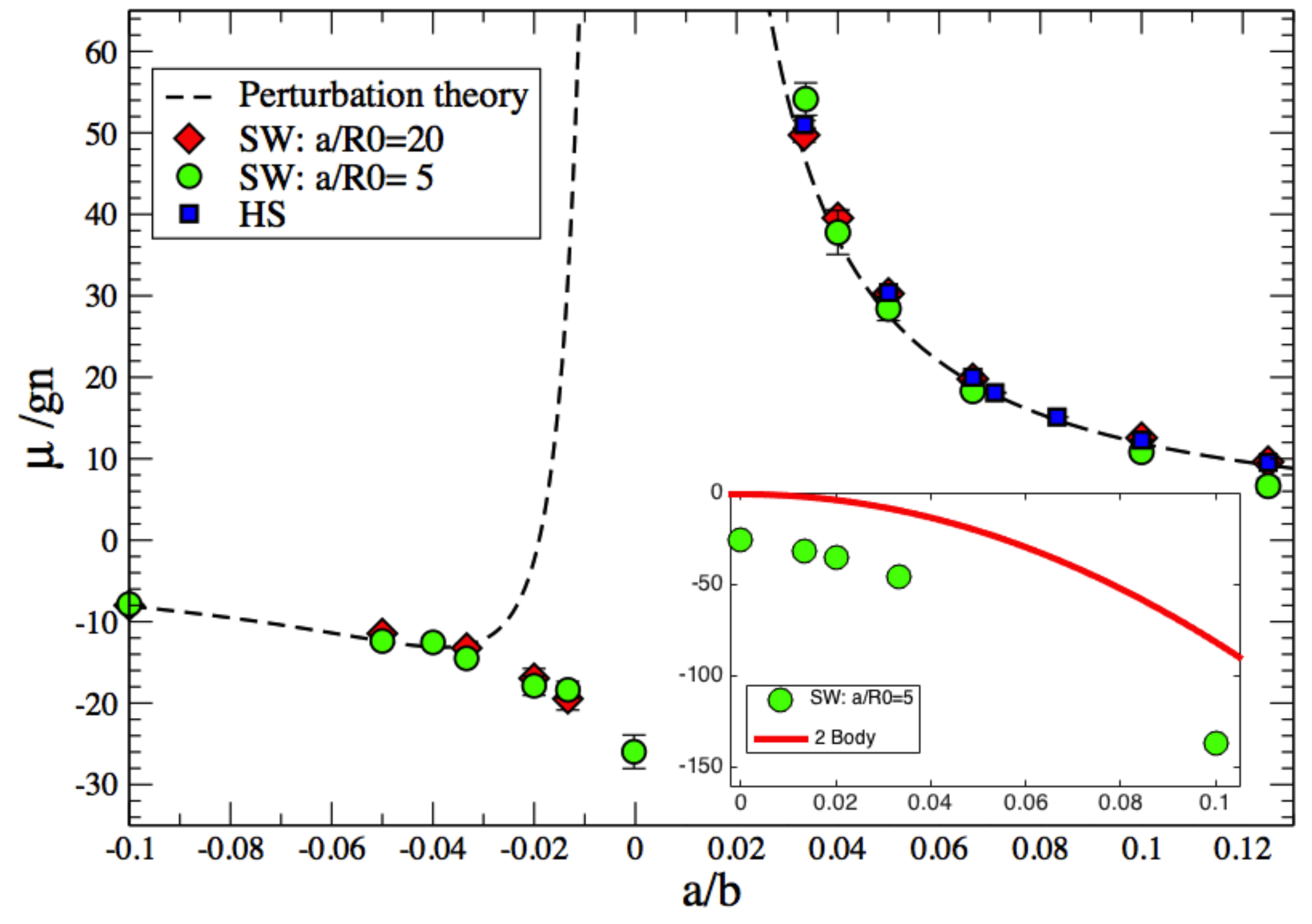}
\caption{(color online). Polaron energy $\mu$ as a function of the ratio $a/b$ of scattering lengths for both the repulsive and the attractive branch. The gas parameter of the bosonic bath is $na^3=10^{-5}$. The symbols are the DMC results obtained with the following impurity-boson interaction potential: hard sphere (blue squares); square well with $a/R_0=5$ (green circles); square well with $a/R_0=20$ (red diamonds). The dashed line is the result (\ref{Bogoliubov7}) of perturbation theory for the two branches. Inset: Polaron energy along the attractive branch on the positive side of the resonance value for the impurity-boson scattering length. The solid line corresponds to the binding energy $\epsilon_b$ in the square well potential with $a/R_0=5$.}
\label{fig1}
\end{center}
\end{figure}

The results for both branches are shown in Fig.~\ref{fig1}. The gas parameter of the bosonic bath 
is here $na^3=10^{-5}$ corresponding to a dilute gas whose ground-state energy $E_0(N)$ is found to be very close to the result (\ref{Bogoliubov2}) of second order perturbation theory. The results reported in Fig.~\ref{fig1} are obtained both with the hard-sphere potential (\ref{eq:Vimp1})
and with the square-well potential (\ref{eq:Vimp2}) where we used two different values of the ratio $a/R_0$ of the boson-boson scattering length to the potential range. The figure clearly indicates that the energies scale with the ratio $a/b$ and that the details of the impurity-boson potential are irrelevant. For the repulsive branch we find a remarkably good agreement with the perturbation result in Eq.~(\ref{Bogoliubov7}) up to values of $b\simeq25a$. For larger values of the impurity-boson scattering length calculations using the square-well potential get increasingly difficult because of large fluctuations arising from the nodal constraint imposed on the Jastrow correlation term $f_I$ which is no longer adequate to define the excited state of the polaron. It is worth stressing at this point that the results for the repulsive branch strongly depend on the choice made for the nodal constraint which provides the correct description of the repulsive polaron only in the limit $nb^3\ll1$. The attractive branch can instead be followed, starting from small negative values of $b$, down to the unitary point ($a/b=0$) and when approaching this point the polaron binding energy shows large deviations from the perturbation expansion (\ref{Bogoliubov7}). We find that at the unitary point $|\mu|\simeq25 gn$, resulting in a binding energy of the impurity  much larger than the chemical potential of bosons in the bath. The results corresponding to positive values of $b$ along the attractive branch are reported in the inset of Fig.~\ref{fig1}). We find that $\mu$ lies always significantly below the two-body binding energy  $\epsilon_b$. One expects that, by increasing $a/b$ on the positive side of the resonance, the polaron binding energy eventually approaches the energy of the deepest cluster state (see Sec.~III).

\begin{figure}
\begin{center}
\includegraphics[height=6.5cm]{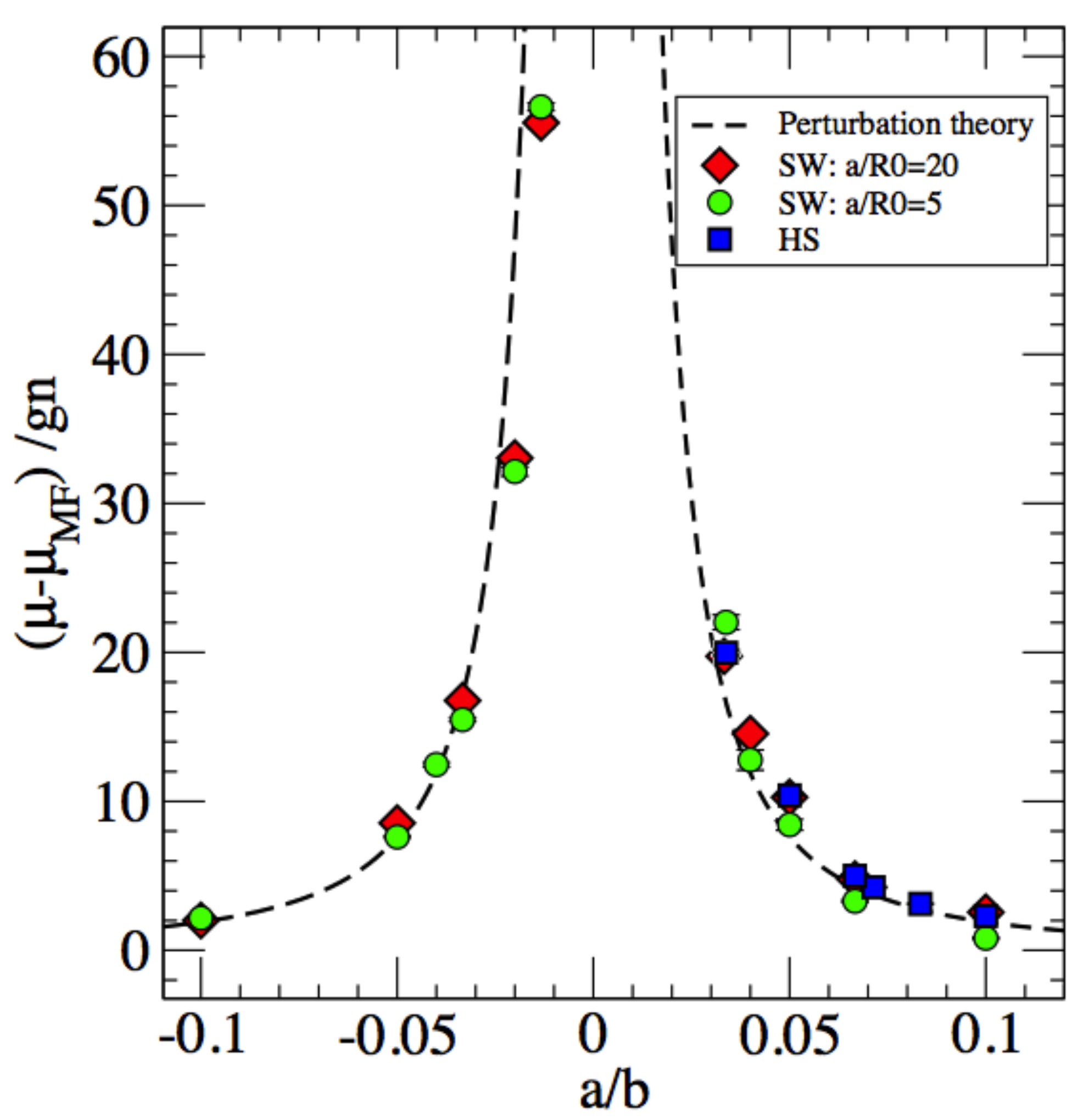}
\caption{(color online). Polaron energy $\mu$ with the mean-filed contribution $\mu_{MF}$ subtracted as a function of the ratio $a/b$. The symbols are as in Fig.~\ref{fig1}. The dashed line is the second order contribution to the perturbation expansion (\ref{Bogoliubov7}) (second term in bracket). Notice that the results for $a/b>0$ refer to the repulsive branch only.}
\label{fig2}
\end{center}
\end{figure}

In Fig.~\ref{fig2} we show a more detailed comparison of the binding energy of the two polaron branches with the perturbation expansion (\ref{Bogoliubov7}). This is carried out by subtracting from the values of $\mu$ the mean-field contribution $\mu_{MF}=gn\frac{b}{a}$, {\it i.e.} the first term in bracket in Eq.~(\ref{Bogoliubov7}), and by comparing $\mu-\mu_{MF}$ with the second order contribution of the perturbation expansion. We notice that $\mu-\mu_{MF}$ remains positive in the limit $a/|b|\ll1$, in contrast with the predictions of Ref.~\cite{Cucchietti06} and \cite{Tempere09} where this quantity should turn negative for $a/|b|\lesssim0.07$ as a consequence of the self-localzation of the polaron.

\begin{itemize}[leftmargin=*]
\item{\underline{Effective mass}}
\end{itemize}

\begin{figure}
\begin{center}
\includegraphics[width=7.5cm]{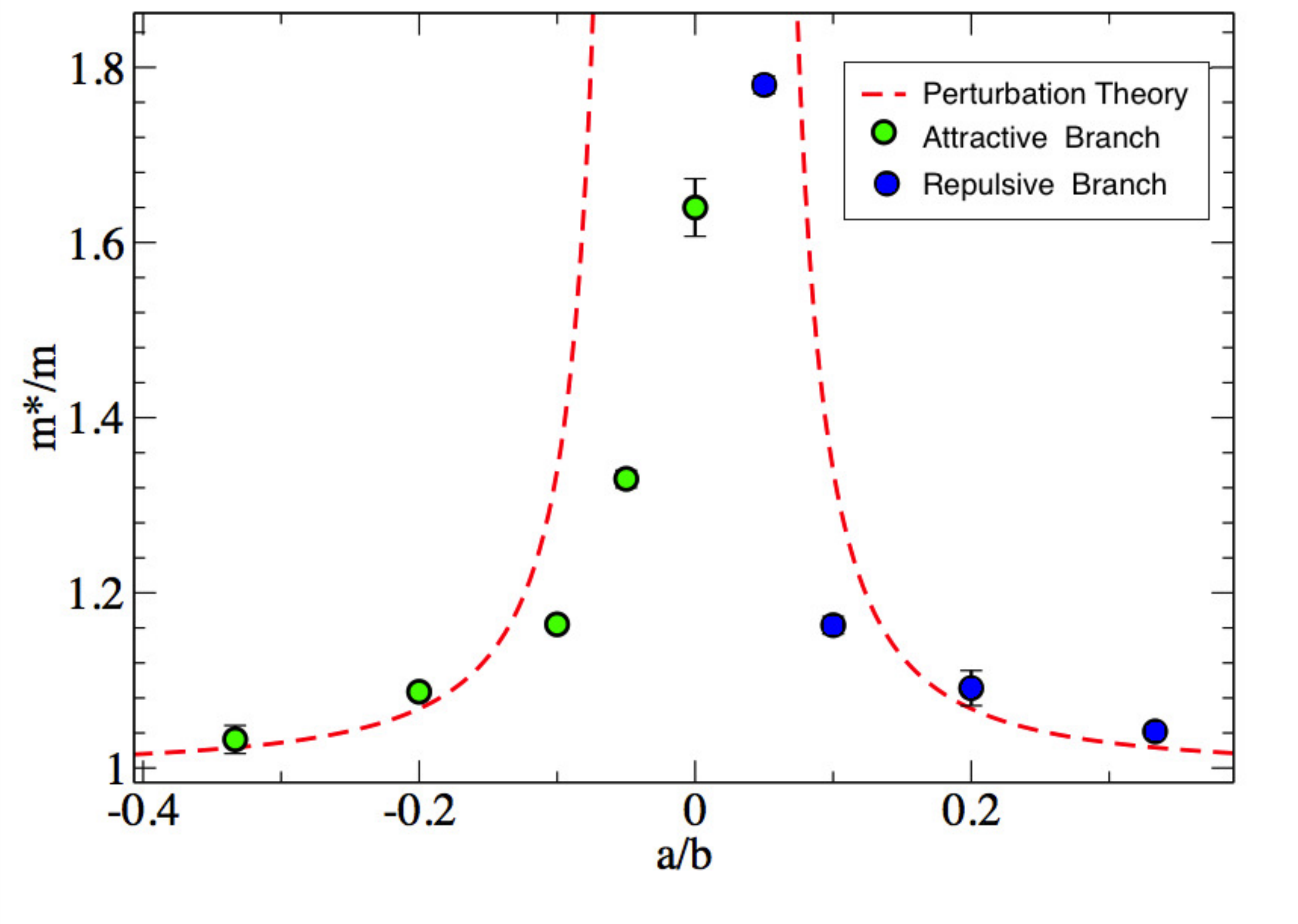}
\caption{(color online). Effective mass of the polaron as a function of the ratio $a/b$ for the repulsive (blue circles) and attractive (green circles) branch. The gas parameter is given by $na^3=10^{-5}$. The dashed line corresponds to the perturbation expansion (\ref{Bogoliubov8}).}
\label{fig3}
\end{center}
\end{figure}

The effective mass of a distinguishable particle can be determined in a DMC simulation by calculating its diffusion constant in imaginary time~\cite{Boninsegni95,Boronat99}. The main assumption is that the energy of the system with the impurity having momentum ${\bf p}_\alpha$ can be written in the form
\begin{equation}
E(N,1;{\bf p}_\alpha)=E_0(N)+\mu+\frac{p_{\alpha}^{2}}{2m^{*}}+\cdots
\label{emass1}
\end{equation}
in terms of the impurity binding energy $\mu$ and effective mass $m^\ast$. The ratio $m/m^\ast$ of the bare to the effective mass of the particle is then given by
\begin{equation}
\frac{m}{m^\ast}=\lim_{\tau\rightarrow\infty}\frac{\langle \left|\Delta{\bf r}_{a}(\tau)\right|^{2}\rangle }{6D\tau} \;,
\label{emass2}
\end{equation}
where $D=\hbar^2/2m$ is the diffusion constant of a free particle and $\langle \left|\Delta{\bf r}_{a}(\tau)\right|^{2}\rangle =\langle \left|{\bf r}_{a}(\tau)-{\bf r}_{a}(0)\right|^{2}\rangle $ is the mean square displacement of the impurity in imaginary time. One can determine the value of $m/m^\ast$ from the large time slope of $\langle \left|\Delta{\bf r}_{a}(\tau)\right|^{2}\rangle$ as a function of the imaginary time $\tau$. The results are shown in Fig.~\ref{fig3} for the attractive and repulsive branch of the polaron. We notice that far away from the resonant point $a/b=0$ the increase of the effective mass agrees with the prediction of perturbation theory. On approaching the resonance, $m^\ast/m$ remains finite reaching values $\lesssim2$. This result is again in contrast with the self-localization picture of Refs.~\cite{Cucchietti06,Tempere09} which predicted a large increase of the effective mass with increasing coupling strength. Along the attractive branch we calculated $m^\ast$ only up to the unitary point, we expect that following this branch on the positive side of the resonance the value of $m^\ast$ should continue to increase.

\begin{itemize}[leftmargin=*]
\item{\underline{Density profiles}}
\end{itemize}

Another important output of our QMC simulations, useful to understand the changes induced in the bosonic bath by the impurity, is the pair correlation function $g_I(r)$ giving the probability of finding a bosonic particle at a distance $r$ from the impurity. At large distances $g_I(r)\simeq1$, whereas its short-range behavior is determined by the impurity-boson potential $V_I(r)$. The density profile of the particles of the bath surrounding the impurity can be calculated using the following integral of the pair correlation function
\begin{equation}
n(r)=n\frac{\int_0^r dr^\prime {r^\prime}^2 g_I(r^\prime)}{r^3/3} \;,
\label{densityprofile1}
\end{equation}
which approaches the bulk value $n$ far away from the impurity. As a technical remark, we compute the function $g_I(r)$ by carrying out both a variational and a diffusion Monte Carlo calculation, which provide the estimates $g_I^{VMC}$ and $g_I^{DMC}$ respectively, and by using the extrapolation formula $g_I(r)=[g_I^{DMC}(r)]^2/g_I^{VMC}(r)$~\cite{Kolorenc11}.

\begin{figure}
\begin{center}
\includegraphics[angle=-90,width=9.0cm]{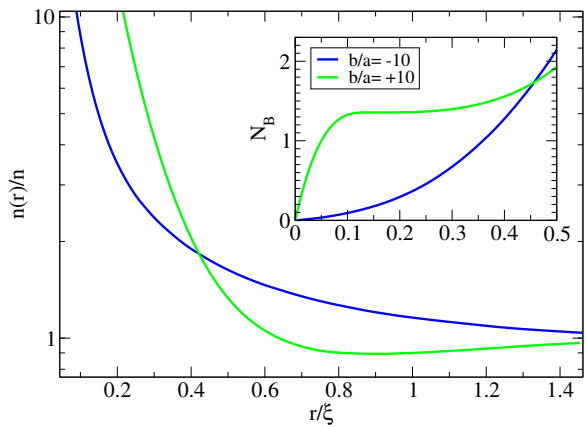}
\caption{(color online). Density profile of the bath surrounding the impurity for $b/a=\pm 10$. Inset: Integrated number of particles of the bath at a distance $r$ from the impurity.}
\label{fig4}
\end{center}
\end{figure}

\begin{figure}
\begin{center}
\includegraphics[angle=-90,width=8.5cm]{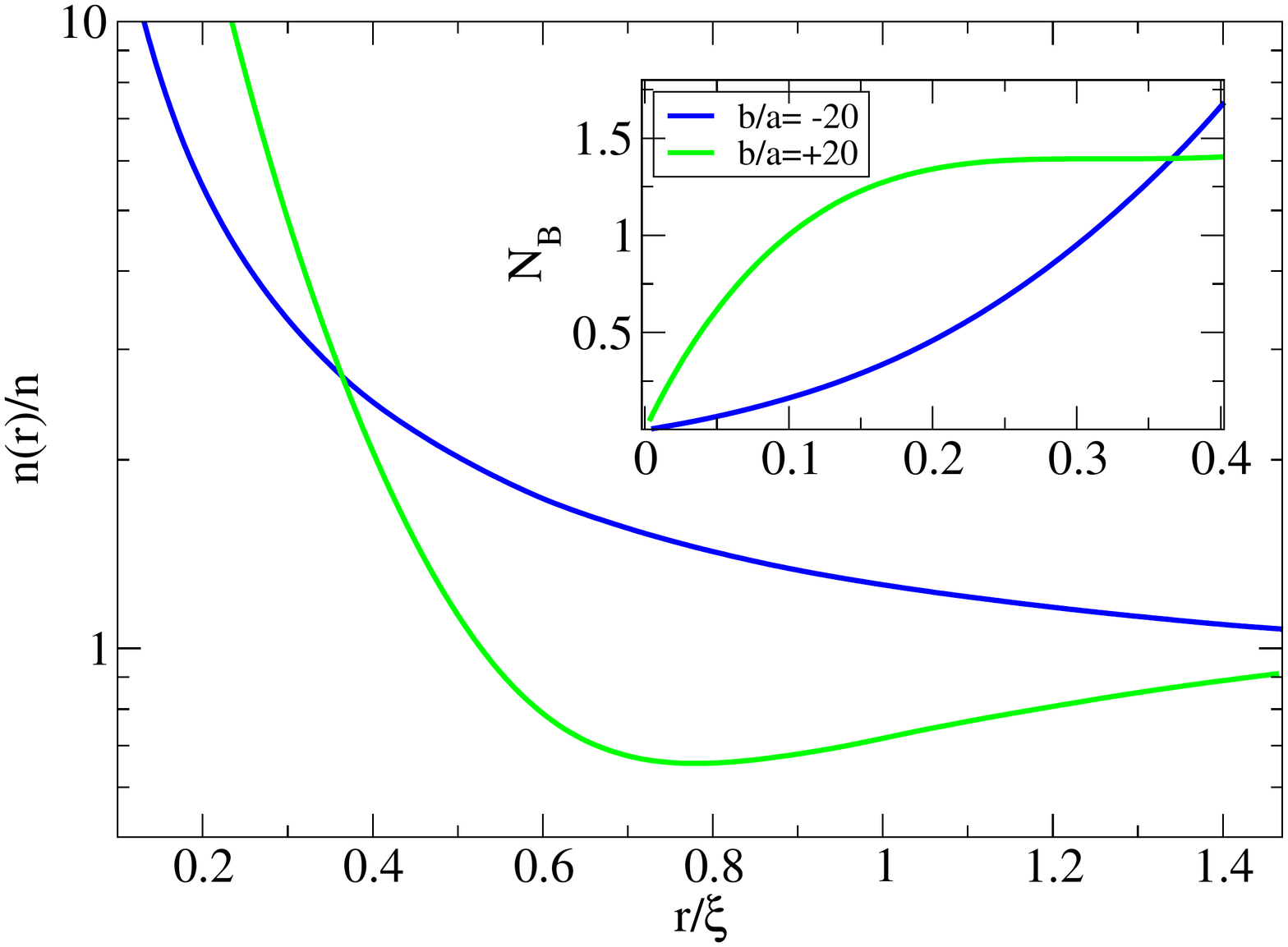}
\caption{(color online). Density profile of the bath surrounding the impurity for $b/a=\pm 20$. Inset: Integrated number of particles of the bath at a distance $r$ from the impurity.}
\label{fig5}
\end{center}
\end{figure}

\begin{figure}
\begin{center}
\includegraphics[angle=-90,width=8.5cm]{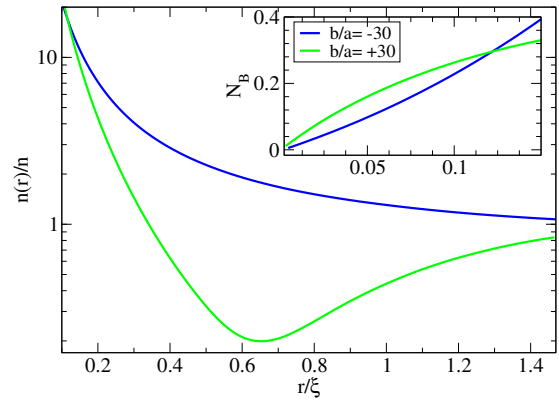}
\caption{(color online). Density profile of the bath surrounding the impurity for $b/a=\pm 30$. Inset: Integrated number of particles of the bath at a distance $r$ from the impurity.}
\label{fig6}
\end{center}
\end{figure}

The ratio $n(r)/n$ of the local to the bulk density is shown in Figs.~\ref{fig4}-\ref{fig6} for three values of $|b|/a$, both on the attractive and the repulsive branch of the polaron. The calculations are carried out at the bath density $na^3=10^{-5}$, using the square-well potential (\ref{eq:Vimp2}) with $a/R_0=5$, and distances are reported in units of the healing length $\xi=1/\sqrt{8\pi na}$. In all cases $n(r)$ exhibits a pronounced peak at the position of the impurity caused by the attractive potential well. If $b<0$ (attractive branch) the local density of particles decreases monotonously, reaching the bulk value $n$ when $r/\xi\gtrsim1$. Instead, if $b>0$ (repulsive branch) the density $n(r)$ goes through a minimum before reaching the bulk value. The position of the minimum lies in the region $0.6\lesssim r/\xi\lesssim0.8$ and decreases with increasing $b/a$. In this case, the density depletion occurring at a large distance from the impurity arises from the effective repulsive interaction associated with the positive value of $b$. 

The average number of particles of the bath surrounding the impurity is obtained from the integral $N_B=4\pi n\int_0^r dr^\prime {r^\prime}^2 g_I(r^\prime)$ and is shown in the inset of Figs.~\ref{fig4}-\ref{fig6} as a function of the distance $r$. We notice that the number $N_B$ starts to grow faster with increasing $r$ for positive values of $b$, consistently with the larger peak of $n(r)$ at very short distances. In particular, for $b/a=10$ and 20 (see Figs.~\ref{fig4}-\ref{fig5}), $N_B$ rapidly reaches the value of one particle already at distances $r/\xi\sim0.1$. 

We also notice that any perturbative approach based on the Fr\"ohlich-type Hamiltonian (\ref{Bogoliubov6}) can only be meaningfully applied if the density perturbation induced by the impurity, $\delta n(r)=(n(r)-n)$, satisfies the condition $|\delta n(r)|/n\ll1$~\cite{Demler14-1}. These approaches are therefore limited to values of $r$ such that $r/\xi\gtrsim1$ and can never describe correctly the structural properties of the bath at short distances from the impurity.

\begin{itemize}[leftmargin=*]
\item{\underline{Contact parameter}}
\end{itemize}

\begin{figure}
\begin{center}
\includegraphics[angle=-90,width=8.5cm]{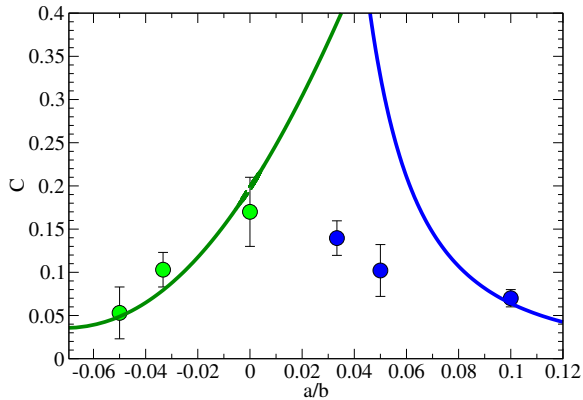}
\caption{(color online). Contact parameter Eq.~(\ref{contact1}) as a function of the ratio $a/b$ for the repulsive (blue circles) and attractive (green circles) branch. The gas parameter is given by $na^3=10^{-5}$. The lines correspond to the determination of $C$ from the attractive and repulsive branch of the equation of state [see Eq.~(\ref{contact2})].}
\label{fig7}
\end{center}
\end{figure}

The impurity-boson contact parameter can be determined from the behavior of the pair correlation function $g_I(r)$ in the range of distances $r\ll n^{-1/3}$, but still much larger than the typical radius of the impurity-boson interaction. We define the dimensionless contact parameter as
\begin{equation}
C=\lim_{r\to0}g_I(r)\frac{r^2}{a^2}(na^3)^{2/3} \;,
\label{contact1}
\end{equation}
where the $r\to0$ limit should be intended in the sense specified above. The results for the contact parameter, obtained at $na^3=10^{-5}$ using the SW potential with $a/R_0=5$, are shown in Fig.~\ref{fig7} for both the attractive and the repulsive branch. For small values of $|b|/a$ the impurity-boson pair correlation function is well approximated by the simple expression $g_I(r)=(1-\frac{b}{r})^2$, determined solely by two-body physics, yielding the estimate $C=(na^3)^{2/3}\frac{b^2}{a^2}$ for the contact parameter. The derivative of the polaron binding energy with respect to the inverse scattering length $b$ should also be related to $C$~\cite{Werner12}. From the behavior in the $|b|/a\ll1$ regime one finds
\begin{equation}
C=\frac{(na^3)^{2/3}}{gn}\frac{d\mu}{d(-a/b)} \;.
\label{contact2}
\end{equation} 
This result is shown in Fig.~\ref{fig7} together with the contact extracted from the pair correlation function. Good agreement is found along the attractive branch, whereas the two estimates of $C$ on the repulsive branch are compatible only in the weak-coupling limit. The disagreement between the contact parameter obtained from the equation of state and from the pair correlation function indicates that our choice of the trial wave function does not provide a fully satisfactory description of the repulsive polaron in the region where $b/a$ becomes very large.

\subsection{E. Resonant interaction}

\begin{figure}
\begin{center}
\includegraphics[angle=-90,width=9.0cm]{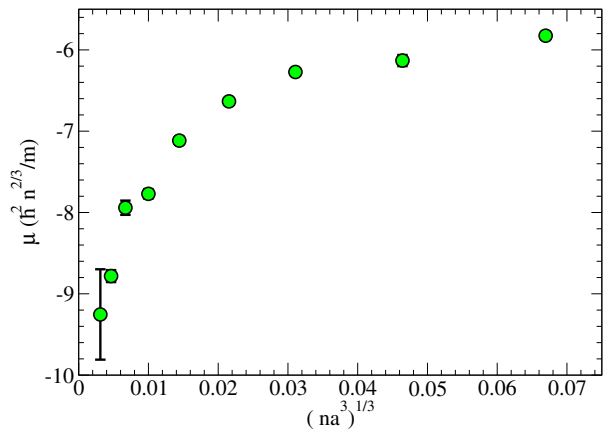}
\caption{Polaron binding energy at unitarity ($a/b=0$) in units of $\frac{\hbar^2n^{2/3}}{m}$ as a function of the gas parameter of the bath.}
\label{fig8}
\end{center}
\end{figure}

In this section we focus on the properties of the Bose polaron when the interaction between the impurity and the bath is resonant, {\it i.e.} $a/b=0$. In Fig.~\ref{fig8} we show the binding energy of the polaron calculated at resonance as a function of the gas parameter of the bath. The results show that $\mu$ scales with the energy $\frac{\hbar^2n^{2/3}}{m}$ and that, once expressed in these units, it depends weakly on the gas parameter over many orders of magnitude. As $na^3$ decreases the value of $\mu$ also decreases, reaching $\mu\simeq-9\frac{\hbar^2n^{2/3}}{m}$ at the very small density $na^3=3\times10^{-8}$. We can not establish weather the binding energy continues to decrease for even smaller densities, signaling the instability of the non interacting gas in the presence of an impurity with attractive interaction, or it reaches a constant value in agreement with the findings of the field-theoretical calculation in Ref.~\cite{Rath13}. Remarkably, the binding energy of a Fermi polaron resonantly interacting with the bath is given by $\mu=-4.4\frac{\hbar^2n^{2/3}}{m}$, where $n$ is here the density of the Fermi sea~\cite{Lobo06}, and differs approximately by a factor of two compared to the results in Fig.~\ref{fig8} for the smallest values of $na^3$. 

\begin{figure}
\begin{center}
\includegraphics[width=9.0cm]{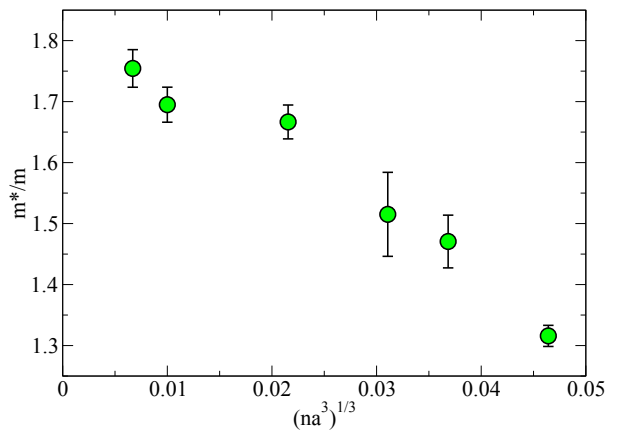}
\caption{Effective mass of the polaron at unitarity ($a/b=0$) as a function of the gas parameter of the bath.}
\label{fig9}
\end{center}
\end{figure}

The effective mass as a function of the gas parameter is shown in Fig.~\ref{fig9}. Also in this case we find a small variation of $m^\ast/m$ following a change of $na^3$ over orders of magnitude. The largest effective mass, $m^\ast/m\simeq1.7$, is achieved at the smallest densities of the bath.

\begin{figure}
\begin{center}
\includegraphics[width=9.0cm]{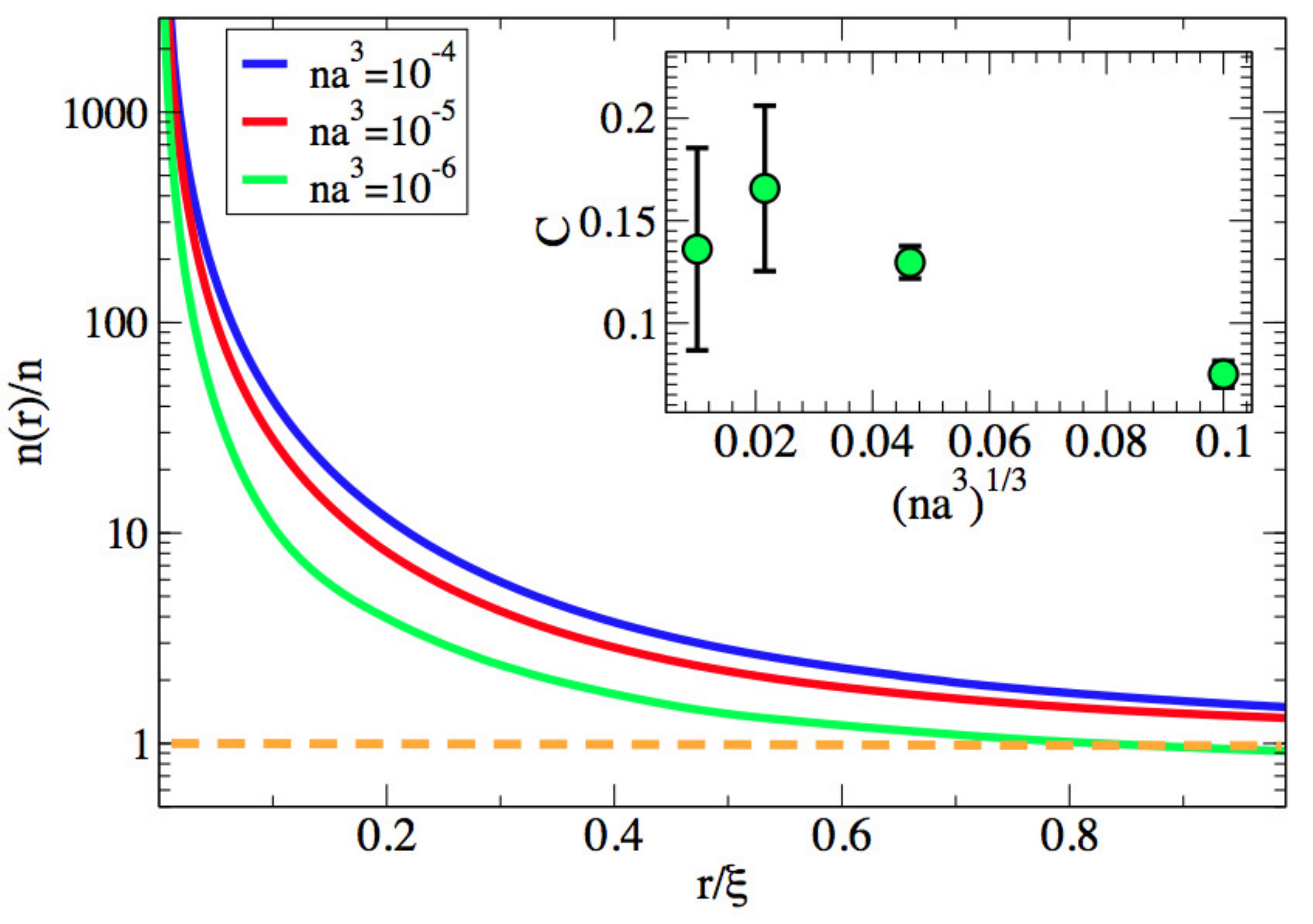}
\caption{Density profile of the bath surrounding the impurity at unitarity ($a/b=0$) for three values of the gas parameter. Inset: Contact parameter $C$ at unitarity as a function of $(na^3)^{1/3}$.}
\label{fig10}
\end{center}
\end{figure}

In Fig.~\ref{fig10} we show the density profile of the bath surrounding the impurity obtained using 
the pair correlation function $g_I$ and Eq.~(\ref{densityprofile1}). The behavior is qualitatively similar to the one reported in Fig.~\ref{fig6} and corresponding to $b/a=-30$ along the attractive polaron branch. By decreasing the value of the bath gas parameter we find that the density peak around the impurity sharpens and the size of the deformation in units of the healing length shrinks. Finally, in the inset of Fig.~\ref{fig10}, we show the value of the contact parameter $C$, determined from the short-range behavior of the pair correlation function, for different values of $(na^3)^{1/3}$. Also for this quantity  we observe a weak dependence on the value of the bath gas parameter.

\section{III. Few-body physics}

In this Section we consider the problem of the existence of bound states in vacuum consisting of the impurity and a number $N$ of bosons. Of course, such bound states can only occur in the case of the square-well model for the impurity-boson potential. This potential supports a two-body molecular state having energy $\epsilon_b$, given by Eq.~(\ref{eq:Vimp4}), for all positive values of the inter-species scattering length $b$.

\begin{figure}
\begin{center}
\includegraphics[width=9.0cm]{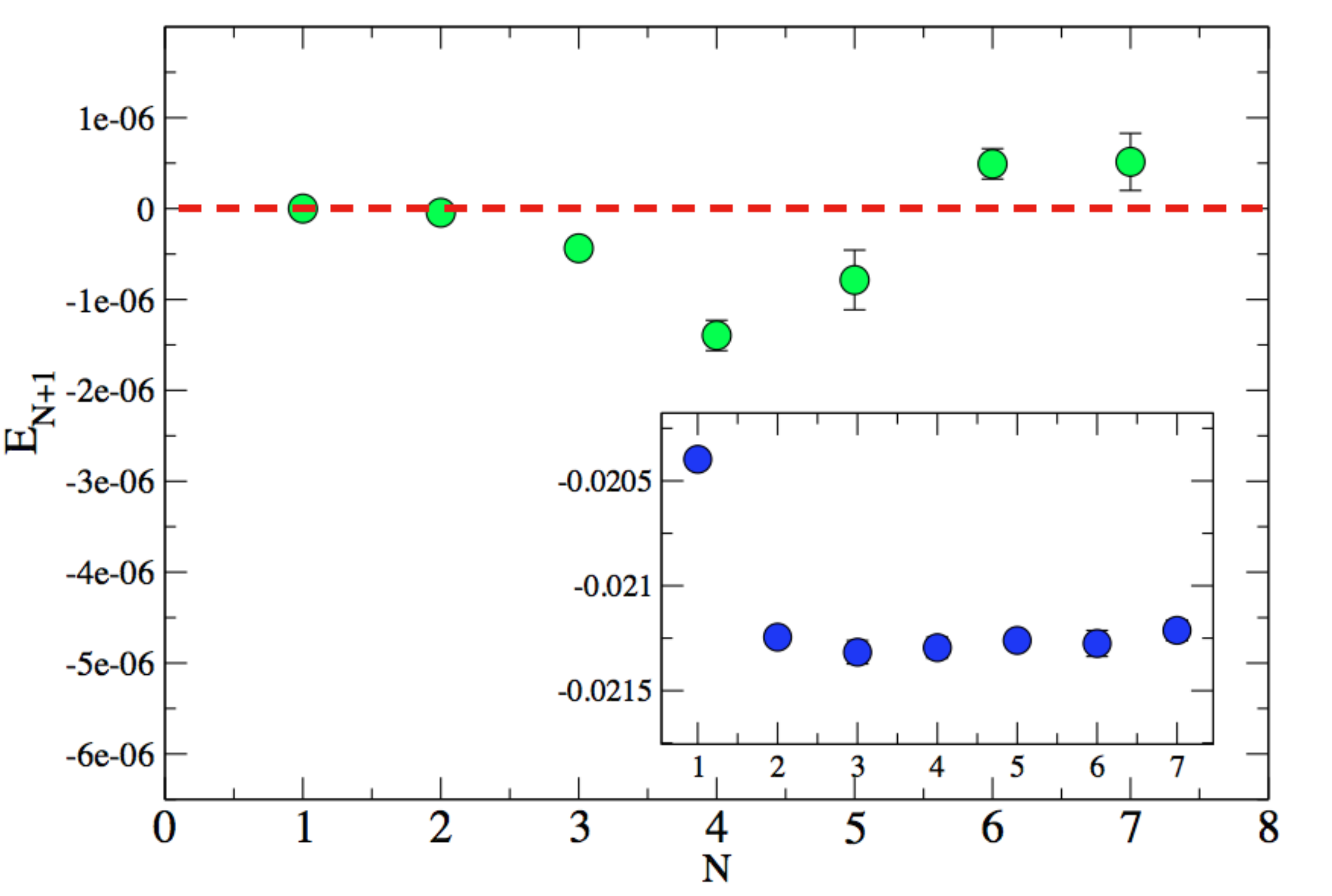}
\caption{Energy of clusters with $N+1$ particles at the unitary point ($a/b=0$) as a function of the number $N$ of bosons. Energies are in units of $\frac{\hbar^2}{2ma^2}$. The value at $N=1$ refers to the two-body binding energy. Inset: Same as main figure at $a/b=0.1$.}
\label{fig10-A}
\end{center}
\end{figure}

The search for the ground state of clusters with $N+1$ particles is carried out using the DMC method based on the following trial wave function    
\begin{equation}
\psi_T({\bf R})=\exp(-\beta R_H) \; \prod_{i=1}^N f_I(r_{i\alpha})\prod_{i<j}f_B(r_{ij}) \;.
\label{FB1}
\end{equation}
The above wave function differs from the one of  Eq.~(\ref{trial1}), used in simulations of homogeneous configurations, by the exponential term which depends on the hyper-radius of the cluster
\begin{equation}
R_H=\sqrt{({\bf r}_\alpha-{\bf r}_{CM})^2+\sum_{i=1}^N({\bf r}_i-{\bf r}_{CM})^2} \;,
\label{FB2}
\end{equation} 
where ${\bf r}_{CM}=\frac{1}{N+1}\left({\bf r}_\alpha +\sum_{i=1}^N{\bf r}_i\right)$ is the coordinate of the center of mass. The Jastrow correlation terms in Eq.~(\ref{FB1}) are similar to the ones of Eqs.~(\ref{Jastrow1}) and (\ref{Jastrow4}), respectively for the boson-boson $f_B$ and the impurity-boson $f_I$ function. Since periodic boundary conditions are absent here, the length scale $L/2$ in Eq.~(\ref{Jastrow1}) is replaced by the large distance $\bar{R}_L=300a$. Moreover, the boundary condition on the derivative of $f_I$ is relaxed with the choice $f_I(r)=B+Ce^{-\alpha r}$, holding for $r>\bar{R}$, with the constants $B$ and $C$ determined in the same way as in Eq.~(\ref{Jastrow4}). Free parameters that are optimized using a variational procedure are the matching point $\bar{R}$ and the coefficients $\alpha$ and $\beta$. In particular, the latter  fixes the size of the cluster in terms of its hyper-radius. 

Calculations are performed in the reference frame where ${\bf r}_{CM}=0$, in order to eliminate the contribution from the center of mass motion. Furthermore, we consider only the resonant point where $a/b=0$ and $\epsilon_b=0$, and the point $a/b=0.1$ on the positive side of the resonance where $\epsilon_b<0$. In Fig.~\ref{fig10-A} we show the results for the ground-state energy of the cluster with $N+1$ particles as a function of the number $N$ of bosons. At unitarity the two-body binding energy, corresponding to $N=1$ in Fig.~\ref{fig10-A}, is identically zero, whereas the three-body Efimov state ($N=2$) is found to feature an extremely shallow ground-state energy: $|E_{2+1}|\lesssim10^{-7}\frac{\hbar^2}{2ma^2}$. This result is consistent with the prediction $E_{2+1}=-\frac{\hbar^2}{mr_0^2}4e^{-2\pi/s_0}$ for the lowest Efimov state in terms of the three-body length $r_0$ and the Efimov parameter $s_0$~\cite{Petrov10}. In the case of equal masses for the impurity and the bosons, the value of $s_0$ is very small, $s_0=0.4137$, resulting in $E_{2+1}\sim-10^{-6}\frac{\hbar^2}{mr_0^2}$ which is of the same order as our estimate if $r_0\sim a$. For increasing $N$ the cluster ground-state energy decreases markedly up to $N=4$, while clusters with $N=6$ are undoubtedly unbound. These findings are compatible with the results at $a/b=0.1$ (see inset of Fig.~\ref{fig10-A}), where the binding energy appears not to decrease further already for $N>3$.

It is important to stress that the largest size $N$ of bound clusters, as well as the precise value of their ground-state energies, depend on the details of the inter-boson and impurity-boson interactions. However, we believe that the qualitative behavior emerging from our simulations should hold for any short-range interaction with scattering length $a$ and $b$, respectively. In particular, we notice that the polaron binding energy shown in Fig.~\ref{fig1} is more than a factor $10^3$ larger than the deepest cluster state at unitarity and remains larger also at $a/b=0.1$. A possible reason for the irrelevance of cluster states at unitarity is the feature of equal masses for the impurity and the bosons, which makes Efimov states extremely shallow.

\section{IV. Many impurities}

Let us now analyze the case of a small concentration of impurities immersed in a BEC at $T=0$. For this problem the statistics of the impurities is important and in the present study we consider only impurities which obey Bose statistics. A binary Bose-Fermi mixture with a small concentration of bosons in a Fermi sea and featuring resonant Bose-Fermi interactions has been investigated using QMC methods in Ref.~\cite{Bertaina13}.   

A collection of $M$ impurities immersed in a gas of $N$ particles is described by the Hamiltonian
\begin{eqnarray}
H&=&-\frac{\hbar^2}{2m_B}\sum_{i=1}^N \nabla_i^2+\sum_{i<j}V_B(r_{ij}) 
\label{Manyimpurities1}\\
&-&\frac{\hbar^2}{2m_I}\sum_{\alpha=1}^M \nabla_\alpha^2
+\sum_{\alpha<\beta}V_{II}(r_{\alpha\beta})
+\sum_{i=1}^N\sum_{\alpha=1}^MV_I(r_{i\alpha}) \;.
\nonumber
\end{eqnarray}
The impurity-impurity potential is modeled by the same hard-sphere interaction, including the same scattering length $a$, which characterizes the coupling between the bosons of the bath: $V_{II}(r)=V_B(r)$. We also assume that the masses of the two types of particles are the same ($m_I=m_B$), and the impurity-boson interaction $V_I(r)$ is as described in Sec.~II-A.

The perturbation treatment of a binary mixture of Bose condensates at $T=0$ has been carried out in Ref.~\cite{Balabanyan86} using an extension of the standard Bogoliubov approach. The result for the ground-state energy in the low concentration limit, $x=M/N\ll1$, is obtained as follows
\begin{eqnarray}
E_0=E_B&+&Ngn\left[ \left(\frac{b}{a}+\frac{32}{3\sqrt{\pi}}\sqrt{na^3}\frac{b^2}{a^2} \right)x\right.
\nonumber\\
&+& \left. \left(1+\frac{64}{3\sqrt{\pi}}\sqrt{na^3}\frac{b^2}{a^2} \right) \frac{x^2}{2} \right] \;,
\label{Manyimpurities2}
\end{eqnarray}   
up to quadratic contributions in the ratio $b/a$ of scattering lengths and in the impurity concentration. Here $E_B$ is the energy (\ref{Bogoliubov2}) of the bath without impurities. Furthermore, one should notice that the term linear in the concentration coincides with the single polaron energy of Eq.~(\ref{Bogoliubov7}), whereas the term proportional to $x^2$ describes the repulsive interaction between polarons.

In QMC simulations we calculate the ground-state energy $E(N,M)$ of the mixture of $N$ bosons plus $M$ impurities making use of the following trial wave function 
\begin{equation}
\psi_T({\bf R})=\prod_{i<j}f_B(r_{ij}) \prod_{\alpha<\beta}f_B(r_{\alpha\beta}) 
\prod_{i=1}^N \prod_{\alpha=1}^M f_I(r_{i\alpha})\;,
\label{Manyimpurities3}
\end{equation}
where the same Jastrow factor $f_B$ of Eq.~(\ref{Jastrow2}) accounts for the repulsive correlations between the particles of the bath and between the impurities while the impurity-boson term $f_I$ is as described in Sec.~II-C. We notice that the above wave function is symmetric under the exchange  of the impurity coordinates fulfilling Bose statistics. 

\begin{figure}
\begin{center}
\includegraphics[width=7.0cm,angle=-90]{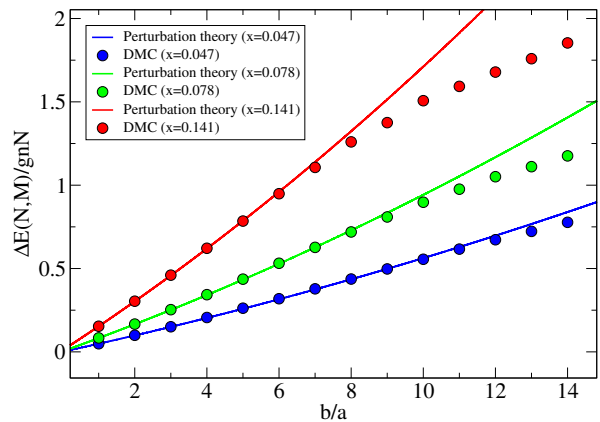}
\caption{Energy shift between the system with a given concentration $x$ of impurities and the bath without impurities as a function of the ratio of scattering lengths along the repulsive branch. The density of the bath is $na^3=10^{-5}$ and results are shown for three different concentrations. Solid lines refer to the perturbation result $E_0-E_B$ from Eq.~(\ref{Manyimpurities2}).}
\label{fig11}
\end{center}
\end{figure}

\begin{figure}
\begin{center}
\includegraphics[width=7.0cm,angle=-90]{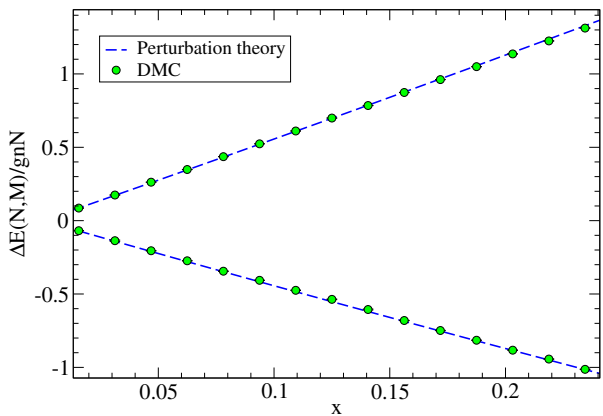}
\caption{Energy shift $\Delta E(M,N)$ as a function of the concentration of impurities for the two values $\frac{b}{a}=\pm5$ of the ratio of scattering lengths. The density of the bath is as in Fig.~\ref{fig11}. Dashed lines refer to the perturbation result $E_0-E_B$ from Eq.~(\ref{Manyimpurities2}).}
\label{fig12}
\end{center}
\end{figure}

We perform QMC calculations using $N=64$ particles in the bath and a varying number $M\le15$ of impurities in a cubic box with periodic boundary conditions, aiming to simulate a homogeneous binary mixture characterized by a small concentration, $x\le0.23$, of the minority component. In particular, we calculate the shift between the ground-state energy of the mixture and of the bath without impurities:  
\begin{equation}
\Delta E(N,M)=E(N,M)-E_0(N) \;.
\label{Manyimpurities4}
\end{equation}
The results for different values of the ratio $b/a$ along the repulsive branch are reported in Fig.~\ref{fig11} for the bath density $na^3=10^{-5}$. The energy shift $\Delta E$ is in remarkable agreement with the perturbation result (\ref{Manyimpurities2}) for all values of $x$ up to $b/a\simeq8$. Only at the largest values of $b/a$ and of the concentration significant deviations from Eq.~(\ref{Manyimpurities2}) are visible. In particular, the interaction between polarons appears to be overestimated by the term proportional to $x^2$ in Eq.~(\ref{Manyimpurities2}). In Fig.~\ref{fig12} we show the results of $\Delta E$ as a function of the concentration $x$ for $b/a=\pm 5$ on both the repulsive and the attractive branch. Also in this case, the comparison with Eq.~(\ref{Manyimpurities2}) shows a very good agreement.

The above results validate the expression (\ref{Manyimpurities2}) for the equation of state of a binary Bose-Bose mixture in the regime of parameters: $|b|/a\lesssim5$ and $x\lesssim0.2$.  Such a validation is important in order to establish the stability conditions and the phase diagram of the mixture in a quantitatively reliable way. From the analysis of the compressibility matrix $\kappa_{ij}=\frac{1}{V}\frac{\partial^2 E(N,M)}{\partial n_i\partial n_j}$, where $n_1=N/V$ and $n_2=M/V$, one finds that the homogeneous mixture is stable if the ratio of scattering length satisfies the condition
\begin{equation}
-1-\frac{104}{3\sqrt{\pi}}\sqrt{na^3}<\frac{b}{a}<1+\frac{8}{3\sqrt{\pi}}\sqrt{na^3} \;.
\label{Manyimpurities5}
\end{equation}
This relation holds in the limit $x\ll1$ and, compared to the mean-field result given by $|b|/a<1$~\cite{Balabanyan86}, includes the leading order correction in the small parameter $\sqrt{na^3}$. We notice that the results reported in Figs.~\ref{fig11} and \ref{fig12} lie outside the stability range (\ref{Manyimpurities5}) of the homogeneous binary mixture. The spinodal instability arising from a vanishing compressibility is associated to systems approaching the thermodynamic limit and is usually prevented in simulations of finite-size systems.

Another possible state of the binary mixture corresponds to a complete phase separation between the $N$ bosons and the $M$ impurities. In this case the ground-state energy can be written as
\begin{eqnarray}
E(n,m)&=&V_1\frac{gn^2}{2}\left(1+\frac{128}{15\sqrt{\pi}}\sqrt{na^3}\right)
\nonumber\\
&+&V_2\frac{gm^2}{2}\left(1+\frac{128}{15\sqrt{\pi}}\sqrt{ma^3}\right) \;,
\label{Manyimpurities6}
\end{eqnarray}
in terms of the densities $n=N/V_1$ and $m=M/V_2$ of the two species and their relative volumes fulfilling the condition $V_1+V_2=V$. The stability of this state requires that $n=m$ and that the energy cost to add one impurity to each of the two uniform phases is positive: $\mu-\frac{dE_0(N)}{dN}>0$. From Eqs.~(\ref{Bogoliubov2}) and (\ref{Bogoliubov7}), giving respectively the energy of the bath without impurities and the excess energy of a single polaron, we get that the phase separated state is stable if 
\begin{align}
\frac{b}{a}&>1   &&  \text{if}\;\; b>0
\label{Manyimpurities7}\\
\frac{|b|}{a}&>\frac{3\sqrt{\pi}}{32\sqrt{na^3}}   && \text{if}\;\; b<0\;.
\label{Manyimpurities8}
\end{align}
One can easily show that, if $b>0$, the energy (\ref{Manyimpurities6}) of the phase separated state lies below the energy (\ref{Manyimpurities2}) of the homogeneous binary mixture for any value $\frac{b}{a}>1$. On the contrary, no gas-like phase appears to be stable outside the regions of Eqs.~(\ref{Manyimpurities7}) and (\ref{Manyimpurities5}) when $b<0$.  

\section{V. Conclusions}

We investigated the properties of an impurity immersed in a Bose-Einstein condensate at $T=0$. 
This Bose polaron study has been carried out using a fully microscopic approach and QMC simulation methods in analogy with previous investigations of the more thoroughly expounded, both theoretically and experimentally, Fermi polaron problem. The main results concern the binding energy of the impurity and its effective mass along the attractive and repulsive polaron branch, explored by changing the scattering length of the impurity-boson interaction potential. 
These results are expected to be universal, for a given boson-boson and impurity-boson scattering length and for equal masses of the two components. At the resonant point of the impurity-boson interaction the polaron binding energy scales with the equivalent of the Fermi energy of the bath in analogy with the behavior found for Fermi polarons. 

The measurement of the polaron binding energy should be accessible in experiments using radio-frequency spectroscopy while its effective mass affects the dispersion of collective excitations in highly imbalanced two-component mixtures. We hope that such experiments, which were so successful in the investigation of the properties of the Fermi polaron, will be carried out also for its Bose counterpart providing a deeper knowledge of this clean and simply stated, but highly non trivial many-body problem.

\section*{Acknowledgments}
We gratefully acknowledge useful discussions with J. Tempere, W. Zwerger, R. Grimm, M. Cetina and D. S. Petrov. This work has been supported by ERC through the QGBE grant.


\begin{thebibliography}{99}
\bibitem{note1} This result was first derived in Ref.~\cite{Saam69} for equal scattering lengths $b=a$ and as a function of the mass ratio $m_I/m_B$.
\end{thebibliography}
\end{document}